\documentclass[12pt]{article}
\usepackage{subeqnarray,indent,amsfonts,amssymb}
\usepackage{bm}    %
\usepackage{graphicx}
\usepackage{url}
\usepackage{hanging}  %

\def\doublespace{ \renewcommand{\baselinestretch}{1.565} \large\normalsize }
\def\singlespace{ \renewcommand{\baselinestretch}{1} \large\normalsize }

\begin{document}

\bibliographystyle{plain}

\date{July 15, 2013 \\[1mm]
    Published in {\em American Psychologist} {\bf 68}, xxx--xxx (2013) \\[1mm]
    \url{http://dx.doi.org/10.1037/a0032850}  \\[2mm]
    \copyright\  2013 American Psychological Association}

\title{\vspace*{-2cm}The Complex Dynamics of Wishful Thinking:\\[1mm]
       The Critical Positivity Ratio}

\author{
   {\normalsize Nicholas J.~L.~Brown}    \\[-1mm]
   {\normalsize\it Strasbourg, France}   \\[4mm]
   {\normalsize Alan D.~Sokal}       \\[-1mm]
   {\normalsize\it New York University and University College London}     \\[4mm]
   {\normalsize Harris L.~Friedman}   \\[-1mm]
   {\normalsize\it Saybrook University and University of Florida}
\vspace*{1cm}
}

\maketitle
\thispagestyle{empty}   %

\singlespace
\begin{center}
\begin{quote}
\begin{quote}
This article may not exactly replicate the final version published in
the APA journal. It is not the copy of record.
\end{quote}
\end{quote}
\end{center}

\bigskip
\noindent
{\bf Running Head:} The critical positivity ratio

\vspace*{1cm}

\begin{center}
{\bf Author note}
\end{center}
\vspace*{-3mm}
\noindent
Nicholas J.~L.~Brown, Strasbourg, France;
Alan D.~Sokal, Department of Physics, New York University
and Department of Mathematics, University College London;
Harris L.~Friedman, Graduate College of Psychology and Humanistic Studies,
  Saybrook University
and Department of Psychology, University of Florida.

We wish to thank Kate Hefferon and Nash Popovic for their encouragement,
and Jean Bricmont, Andrew Gelman, Steven Pinker, and Colin Sparrow
for helpful comments on an early draft of this manuscript.
The authors, of course, bear full and sole responsibility for the
content of this article.

Correspondence concerning this article should be addressed to
Alan D.~Sokal, Department of Physics, New York University,
4 Washington Place, New York, NY 10003.
E-mail: sokal@nyu.edu

\doublespace

\clearpage
\doublespace
\vspace*{0mm}
\begin{center}
{\bf Abstract}
\end{center}
\vspace*{-3mm}
\noindent
We examine critically the claims made by Fredrickson and Losada (2005)
concerning the construct known as the ``positivity ratio.''
We find no theoretical or empirical justification for the use of
differential equations drawn from fluid dynamics, a subfield of physics,
to describe changes in human emotions over time;
furthermore, we demonstrate that the purported application of these equations
contains numerous fundamental conceptual and mathematical errors.
The lack of relevance of these equations
and their incorrect application lead us to conclude that
Fredrickson and Losada's claim to have demonstrated the existence of
a critical minimum positivity ratio of 2.9013 is entirely unfounded.
More generally, we urge future researchers to exercise caution
in the use of advanced mathematical tools such as nonlinear dynamics
and in particular to verify that the elementary conditions
for their valid application have been met.

\bigskip
\bigskip

\clearpage

\newcommand{\apasection}[1]{{\begin{center} {\bf #1} \end{center}\vspace{-3mm}}}
\newcommand{\apasubsection}[1]{{\medskip\par\noindent {\bf #1} \par}}
\newcommand{\apasubsubsection}[1]{{\par\indent {\bf #1.}}}
\newcommand{\apaparagraph}[1]{{\par\indent \textit{\textbf{#1.}} }}

\newcommand{\be}{\begin{equation}}
\newcommand{\ee}{\end{equation}}
\newcommand{\<}{\langle}
\renewcommand{\>}{\rangle}
\newcommand{\widebar}{\overline}
\def\reff#1{(\protect\ref{#1})}
\def\spose#1{\hbox to 0pt{#1\hss}}
\def\ltapprox{\mathrel{\spose{\lower 3pt\hbox{$\mathchar"218$}}
 \raise 2.0pt\hbox{$\mathchar"13C$}}}
\def\gtapprox{\mathrel{\spose{\lower 3pt\hbox{$\mathchar"218$}}
 \raise 2.0pt\hbox{$\mathchar"13E$}}}
\def\textprime{${}^\prime$}

\def\scra{\mathcal{A}}
\def\scrb{\mathcal{B}}
\def\scrc{\mathcal{C}}
\def\scrd{\mathcal{D}}
\def\scre{\mathcal{E}}
\def\scrf{\mathcal{F}}
\def\scrg{\mathcal{G}}
\def\scrl{\mathcal{L}}
\def\scrm{\mathcal{M}}
\def\scro{\mathcal{O}}
\def\scrp{\mathcal{P}}
\def\scrq{\mathcal{Q}}
\def\scrr{\mathcal{R}}
\def\scrs{\mathcal{S}}
\def\scrt{\mathcal{T}}
\def\scrv{\mathcal{V}}
\def\scrw{\mathcal{W}}
\def\scrz{\mathcal{Z}}

\newenvironment{sarray}{
          \textfont0=\scriptfont0
          \scriptfont0=\scriptscriptfont0
          \textfont1=\scriptfont1
          \scriptfont1=\scriptscriptfont1
          \textfont2=\scriptfont2
          \scriptfont2=\scriptscriptfont2
          \textfont3=\scriptfont3
          \scriptfont3=\scriptscriptfont3
        \renewcommand{\arraystretch}{0.7}
        \begin{array}{l}}{\end{array}}

\newenvironment{scarray}{
          \textfont0=\scriptfont0
          \scriptfont0=\scriptscriptfont0
          \textfont1=\scriptfont1
          \scriptfont1=\scriptscriptfont1
          \textfont2=\scriptfont2
          \scriptfont2=\scriptscriptfont2
          \textfont3=\scriptfont3
          \scriptfont3=\scriptscriptfont3
        \renewcommand{\arraystretch}{0.7}
        \begin{array}{c}}{\end{array}}

\pagestyle{myheadings}
\markboth{{\rm Running head: THE CRITICAL POSITIVITY RATIO}}{{\rm Running head: THE CRITICAL POSITIVITY RATIO}}

\doublespace

The ``broaden-and-build'' theory (Fredrickson, 1998, 2001, 2004)
postulates that positive emotions help to develop broad repertoires
of thought and action, which in turn build resilience to buffer against
future emotional setbacks.
Fredrickson and Losada (2005) took the broaden-and-build theory a step
further, by proposing that an individual's degree of flourishing 
could be predicted by that person's ratio of positive to negative
emotions over time, which they termed the ``positivity ratio'' (p.~678).

On its own, the positivity ratio as propounded by Fredrickson and Losada
(2005) is not a particularly controversial construct;
indeed, there is a long history of looking at ratios (e.g., Bales, 1950)
and non-ratio indices (e.g., Bradburn, 1969)
relating positive to negative emotions.
However, Fredrickson and Losada
took matters considerably farther,
claiming to have established that their use of a mathematical model
drawn from nonlinear dynamics provided theoretical support
for the existence of a pair of critical positivity-ratio values
(2.9013 and 11.6346) such that individuals whose ratios fall between
these values will ``flourish,'' while people whose ratios lie outside
this ideal range will ``languish.''  The same article purported to
verify this assertion empirically, by demonstrating that among a group of
college students, those who were ``languishing'' had an average positivity ratio
of 2.3, while those who were ``flourishing'' had an average positivity
ratio of 3.2.

The work of Fredrickson and Losada (2005) has had an extensive influence
on the field of positive psychology.  This article has been frequently cited,
with the Web of Knowledge listing 322 scholarly citations
as of April 25, 2013.
Fredrickson and Kurtz (2011, pp.~41--42), in a recent review,
highlighted this work as providing an ``evidence-based guideline''
for the claim that a specific value of the positivity ratio
acts as a ``tipping point beyond which the full impact of positive
emotions becomes unleashed'' (they now round off 2.9013 to 3).
An entire chapter of Fredrickson's popular book (2009, Chapter~7)
is devoted to expounding this ``huge discovery'' (p.~122),
which has also been enthusiastically brought to a wider audience
by Seligman (e.g., 2011a, pp.~66--68; 2011b).
In fact, the paperback edition of Fredrickson's book (2009) is subtitled
``Top-Notch Research Reveals the 3-to-1 Ratio That Will Change Your Life.''

It is worth stressing that Fredrickson and Losada (2005) did not qualify
their assertions about the critical positivity ratios in any way.
The values 2.9013 and 11.6346 were presented as being independent of
age, gender, ethnicity, educational level, socioeconomic status
or any of the many other factors that one might imagine as potentially
leading to variability.
Indeed, Fredrickson and Losada went so far as to assert (pp.~678, 684, 685)
that the {\em same}\/ critical minimum positivity ratio of 2.9013
applies to individuals, couples, and groups of arbitrary size
(see also Fredrickson, 2009, pp.~133--134).
And yet, the idea that any aspect of human behavior or experience should be universally
and reproducibly constant to five significant digits would, if proven,
constitute a unique moment in the history of the social sciences.
It thus seems opportune to examine carefully the chain of evidence and
reasoning that led
to these
remarkable conclusions.

Fredrickson and Losada (2005) based their assertions concerning the
critical positivity ratios on previous articles by Losada (1999)
and Losada and Heaphy (2004) in which the Lorenz equations
from fluid dynamics were ``applied'' to describe the changes in
human emotions over time.  The major part of the present paper
is devoted, therefore, to a critical examination of the three subject
articles, in chronological order.
We shall demonstrate that each one
of the three articles is completely vitiated by fundamental conceptual
and mathematical errors, and above all by the total absence of any
justification for the applicability of the Lorenz equations to
modeling the time evolution of human
emotions.  (Furthermore,
although the second and third articles rely
entirely
for their validity
on the presumed correctness of their predecessors --- which, as we shall
demonstrate, is
totally
lacking --- we nevertheless invite the reader,
at each stage, to assume for the sake of argument the correctness of the
preceding article(s);  in this way we are able to explain more clearly
the {\em independent}\/ flaws of each of the three subject articles.)
We
conclude that Fredrickson and Losada's (2005) claims
concerning the alleged critical values of the positivity ratio are
entirely unfounded.

\apasection{A brief introduction to differential equations and nonlinear dynamics}

Although the three subject articles rest on a purported application of
differential equations --- a branch of mathematics in which most
psychologists are unlikely to be expert --- to social-psychological data,
none of the three articles provides
its readers with
any more than a vague explanation
of the mathematics on which
those articles
rely.
It therefore seems appropriate to begin this critical analysis
by giving a brief (and we hope pedagogical) introduction to differential
equations, explaining what they are and when they can be validly used.
While we appreciate that it is unusual to find this type of material
in a psychology journal, we nevertheless encourage even the less
mathematically-inclined reader to attempt to follow our explanations
as far as possible, and to consult a mathematician or physicist
friend
for help if needed (and to verify that our arguments are correct).

\apasubsection{What are differential equations?}

Differential equations are employed in the natural and social sciences
to model phenomena in which one or more dependent variables
$x_1,x_2,\ldots,x_n$ evolve deterministically as a function of time ($t$)
in such a way that the rate of change of each variable at each moment
of time is a known function of the values of the variables at that same
moment of time.\footnote{
   More precisely, in this paper we shall be concerned exclusively
   with what mathematicians call {\em ordinary differential equations}\/:
   these are the simplest type of differential equation,
   and are also the type that is employed in the three subject articles.
   For completeness let us mention that there also exist other types of
   differential equations, including
   {\em partial differential equations}\/
   (in which there are two or more independent variables,
    rather than the single independent variable $t$ considered here)
   and {\em stochastic differential equations}\/
   (in which the time evolution involves explicit randomness).
   Some of the statements made here concerning the properties of
   ordinary differential equations require modification
   in connection with these other types of equations.
   For instance, the requirement of deterministic evolution
   (items DE3 and VA3 below) manifestly does not apply to
   stochastic differential equations.
 \label{footnote_ODE}
}

Let us unpack this definition slowly, beginning with the case in which
there is a single dependent variable $x$.

\apasubsubsection{One dependent variable}
We consider a situation in which both the independent variable $t$ (``time'')
and the dependent variable $x$ can be treated as continuous quantities;
we furthermore assume that $x$ varies smoothly as a function of $t$.
First-year calculus then defines the instantaneous rate of change of $x$, conventionally
written $dx/dt$.  A (first-order) {\em differential equation}\/ for the
function $x(t)$ is an equation of the form
\be
   {dx \over dt}   \;=\;   F(x)
 \label{eq.1}
\ee
where $F$ is a {\em known}\/ (i.e., explicitly specified) function.
What this says, in words, is that the rate of change of $x$
at any moment of time is precisely the function $F$ applied to the value of $x$
at that same time.
More specifically, an equation of this form is saying that:

(DE1) Both time ($t$) and the dependent variable ($x$) can be treated
      as continuous quantities.

(DE2) $x$ changes smoothly over time (i.e., it does not jump).

(DE3) $x$ evolves deterministically (i.e., there is no randomness).

(DE4) The rate of change of $x$ at any given moment of time depends
      only on the value of $x$ itself (i.e., not on some additional variables),
      and only on the value of $x$ at {\em that same}\/ moment of time
      (i.e., not on the values in the past).

(DE5) The rate of change of $x$ at time $t$ is exactly $F(x(t))$,
      where $F$ is an explicitly specified function.

The mathematical model of the phenomenon under study thus consists of
Equation~\ref{eq.1} together with the specification of the function $F$.

\apaparagraph{Example 1}
Consider a bank account with continuously compounded interest.
Then the amount of money in the account ($x$) increases with time ($t$)
according to the differential equation
\be
   {dx \over dt}   \;=\;  rx
 \label{eq.2}
\ee
where $r$ is the interest rate.
This has the form of Equation~\ref{eq.1}
with $F(x) = rx$.  This same equation describes the cooling of a coffee cup,
the decay of radioactive atoms, and a vast number of other physical, biological,
and social phenomena.  (In some of these applications, $r$ is a negative
number.)  Equation~\ref{eq.2} is an example of a {\em linear}\/
differential equation, because the function $F(x) = rx$ is linear
(i.e., doubling $x$ causes $F(x)$ to precisely double).

Equation~\ref{eq.2} happens to have a simple solution,
namely $x(t) = x_0 \, e^{rt}$,
where $x_0$ is the account balance at time~0,
and $e \approx 2.71828$ is the base of natural logarithms.
This formula illustrates an important general principle:
the solution of a differential equation
is completely determined by the {\em initial conditions}\/,
that is, the value(s) of the dependent variable(s) at time~0.

It should be stressed that most differential equations
do not have simple solutions that can be explicitly written down;
rather, the solutions have to be studied numerically, by computer.
Nevertheless, the solution at arbitrary time $t$ is always determined,
at least in principle, from the initial conditions,
even though it may not be given by any simple explicit formula.

\apaparagraph{Example 2}
Consider a population $x$ of some species living in a limited territory,
with a maximum sustainable population $X_{\rm max}$.
Then a plausible (though of course extremely oversimplified) model
for the growth of this population is given by the differential equation
\be
   {dx \over dt}   \;=\;  rx \Bigl( 1 \,-\, {x \over X_{\rm max}} \Bigr)
 \label{eq.3}
\ee
where $r$ is some positive number,
which has the form of Equation~\ref{eq.1}
with $F(x) = rx (1 - x/X_{\rm max})$.
Equation~\ref{eq.3} is an example of a
{\em nonlinear}\/ differential equation,
because here the function $F$ is nonlinear
(i.e., doubling $x$ does {\em not}\/ cause $F(x)$ to precisely double).

Of course, population is not, strictly speaking, a continuous
variable, since there is no such thing as a fractional person.
This would appear to conflict with principle DE1 above.
But if the population $x$ is large (e.g., millions of people),
then only a negligible error is made by treating the population
as if it were a continuous variable.

\apasubsubsection{Several dependent variables}
Let us now consider the case in which we have several dependent variables.
We assume once again that both the independent variable $t$
and the dependent variables $x_1,\ldots,x_n$ can be treated as
continuous quantities,
and that $x_1,\ldots,x_n$ vary smoothly as a function of $t$.
Then a {\em system of}\/ (first-order) {\em differential equations}\/
for the functions $x_1(t), \ldots, x_n(t)$
is a system of equations of the form
\begin{eqnarray}
   {dx_1 \over dt}   & = &   F_1(x_1,\ldots,x_n)  \nonumber \\[-1mm]
                     & \vdots &                  \label{eq.system} \\[-2mm]
   {dx_n \over dt}   & = &   F_n(x_1,\ldots,x_n)  \nonumber
\end{eqnarray}
where $F_1,\ldots,F_n$ are specified functions.
The model is constituted by the system of Equations~\ref{eq.system}
together with the specification of the functions $F_1,\ldots,F_n$.

\apaparagraph{Example 3}
Lorenz (1963), building on work by Saltzman (1962),
introduced the following system of differential equations as a
simplified model of convective flow in fluids:
\begin{subeqnarray}
   {dX \over d\tau}   & = &   -\sigma X + \sigma Y  
        \slabel{eq.lorenz.a} \\[3mm]
   {dY \over d\tau}   & = &   rX - Y - XZ
        \slabel{eq.lorenz.b} \\[3mm]
   {dZ \over d\tau}   & = &   -bZ + XY
        \slabel{eq.lorenz.c}
        \label{eq.lorenz}
\end{subeqnarray}
Here the independent variable $\tau$ is a dimensionless variable that is
proportional to time~$t$.
The dependent variables $X$, $Y$, and $Z$ are also dimensionless,
and represent various aspects of the fluid's motion and its temperature gradients.
(In the solutions studied by Lorenz,
$X$ and $Y$ oscillate between positive and negative values,
while $Z$ stays positive.)
Finally, $\sigma$, $b$, and $r$ are positive dimensionless parameters
that characterize certain properties of the fluid and its
flow; in particular, $r$ measures (roughly speaking)
the strength of the tendency to develop convection.

For future reference, let us explain what is meant by ``dimensionless.''
A quantity is called {\em dimensionful}\/ if its numerical value
depends on an arbitrary choice of units.
For instance, lengths (measured in meters or furlongs or \ldots)
are dimensionful, as are times
and masses.
By contrast, a quantity is called {\em dimensionless}\/
if its numerical value does not depend on a choice of units.
For instance, the {\em ratio}\/ of two lengths is dimensionless,
because the units cancel out when forming the ratio;
likewise for the ratio of two times or of two masses.
Physical laws are best expressed in terms of dimensionless quantities,
since these are independent of the choice of units;
this explains why Lorenz (1963) rescaled his equations as he did.
In particular, Lorenz's variable $\tau$ equals $t/T$, where $T$ is
a particular ``characteristic time'' in the fluid-flow problem.

\clearpage  %
\apasubsection{Nonlinear dynamics and chaos}

The mathematical field of {\em nonlinear dynamics}\/
--- popularly known as {\em chaos theory}\/ ---
is founded on the observation that simple equations can under some circumstances
have extremely complicated solutions.
In particular, fairly simple systems of nonlinear differential equations
can exhibit {\em sensitive dependence to initial conditions}\/:
that is, small changes in the initial conditions can lead to deviations
in the subsequent trajectory that grow exponentially over time.
Such systems, while deterministic in principle,
can be unpredictable in practice beyond a limited window of time;
the behavior can appear random even though it is not.
We refer the reader to Lorenz (1993), Stewart (1997) and Williams (1997)
for sober nontechnical introductions to nonlinear dynamics,
and to Strogatz (1994) and Hilborn (2000)
for introductions presuming some background in undergraduate mathematics and physics.

\apasubsubsection{The Lorenz system, revisited}
The Lorenz system (Equations~\ref{eq.lorenz} above) illustrates nicely
some of the concepts of nonlinear dynamics.
Firstly, this system has a {\em fixed point}\/ at $X=Y=Z=0$:
if the system is started there, it stays there forever.
(Physically, this fixed point corresponds to a fluid at rest.)
For $r<1$ it turns out that this fixed point is {\em stable}\/:
if the system is started {\em near}\/ the fixed point,
it will move {\em towards}\/ the fixed point as time goes on.
For $r>1$  this fixed point is {\em unstable}\/:
if started near the fixed point, the system will move {\em away from}\/ it.

For $r > 1$ there is another pair of fixed points,
at $X = Y = \pm \sqrt{b(r-1)}$, $Z = r-1$.
(Physically, these fixed points correspond to a steady-state convective flow.)
It turns out that these fixed points are stable for $r < r_{\rm crit}$
and unstable for $r > r_{\rm crit}$,
where $r_{\rm crit} = \sigma (\sigma+b+3)/(\sigma-b-1)$
and we assume $\sigma > b+1$.

What happens for $r > r_{\rm crit}$?
Lorenz (1963) investigated the trajectories numerically and found that
they tend to a butterfly-shaped set now known as the {\em Lorenz attractor}\/.
This set is a {\em fractal}\/:  it is neither two-dimensional (a surface)
nor three-dimensional (a volume) but something in-between.
Moreover, the trajectories near the attractor
exhibit sensitive dependence to initial conditions
(i.e., they are {\em chaotic}\/).\footnote{
   Actually, the Lorenz attractor is ``born'' at a value $r_{\rm A}$
   slightly less than $r_{\rm crit}$;
   for $r_{\rm A} < r < r_{\rm crit}$, {\em both}\/ the fixed points
   and the Lorenz attractor are stable (Sparrow, 1982, pp.~31--32).
 \label{footnote_rA}
}

One final remark:  Mathematicians studying the Lorenz system
typically (though not always) fix $\sigma$ and $b$ and vary $r$.
Saltzman (1962) chose the illustrative values $\sigma=10$ and $b=8/3$;
then Lorenz (1963) followed him, as have most (though not all) workers ever since.
But there is nothing magical about these values;
indeed, any other values within a fairly wide range would produce qualitatively
similar behavior (Sparrow, 1982, pp.~179--184).

\apasubsection{When can differential equations validly be applied?}

When used properly, differential equations constitute a powerful tool
for modeling time-dependent phenomena in the natural and social sciences.
But a number of preconditions must be met if such an application is
to be valid.  In order to apply differential equations to a specific 
natural or social system, one must first:

   (VA1)
{\em  Identify and define precisely the variables}\/
that specify the state of the system at a given moment of time.
Per principles DE1 and DE2 (above),
these variables must be continuous (or at least approximately so),
not discrete, and they must evolve smoothly in time
(i.e., without jumps).
These variables need not be directly observable quantities;
they can also be (hypothetical or hidden) latent variables,
which are postulated to affect the observable quantities in specified ways.

   (VA2)
Give reasons why these variables can be assumed to
{\em evolve by themselves}\/, without significant effect from other variables
not taken into account in the model (cf.\ principle DE4).

   (VA3)
Give reasons why these variables can be assumed to
{\em evolve deterministically}\/, at least to a good approximation
(cf.\ principle DE3).

   (VA4)
Give reasons why these variables can be assumed to
{\em evolve according to a differential equation}\/
(at least to a good approximation): 
that is, that the rate of change at each moment of time
depends only on the values of the variables at {\em that same}\/ moment of time
(cf.\ principle DE4).

   (VA5)
{\em Find the specific differential equation}\/
giving (at least approximately) that evolution: 
that is, find the functions $F_1,\ldots,F_n$ and give arguments justifying them
{\em for the specific system under study}\/.
In particular, if the functions $F_1,\ldots,F_n$ involve unspecified constants
(such as the constants $r$ and $X_{\rm max}$ in Examples 1 and 2,
or $\sigma$, $b$, and $r$ in Example 3)
and the behavior of the system depends in a significant way on those constants,
then arguments must be given to {\em justify the particular values}\/
that are given to those constants.

Only after
the conditions VA1--VA5
have been fulfilled does it make sense
to apply the theory of differential equations in general,
or the theory of nonlinear dynamics in particular,
to the natural or social phenomenon under study.

It goes without saying that these conditions are not easy to fulfill;
and they become more difficult to fulfill, the more complex is the system
under study.  Consequently
it is not surprising that most of the valid applications
of nonlinear dynamics have arisen in physics and chemistry, where one
can sometimes find systems that are sufficiently simple and isolated
so that one can (a) identify a small number of relevant variables
that evolve by themselves, and (b) write down the equation describing
(at least to a reasonable degree of approximation) their evolution.
See also Ruelle (1994), Kellert (1995), and Sokal and Bricmont (1998, Chapter~7)
for some pertinent cautions concerning the applicability of nonlinear dynamics
to the modeling of real-world phenomena.

There is also a valid ``shortcut'' approach to modeling using differential
equations:   carry out VA1 strictly, but skip VA2--VA5 and simply
{\em guess}\/ (by whatever means) the functions $F_1, \ldots, F_n$;
then obtain empirical data that are sufficiently powerful
to constitute strong evidence that VA2--VA5 hold for the system under study
(though one may not yet understand why).

To summarize:  In all cases it is necessary to carry out VA1 strictly
and to give evidence that VA2--VA5 hold in one's application;
but there are two alternatives concerning the type of evidence one supplies.
In the ``standard'' approach one gives theoretical arguments
that VA2--VA5 hold,
while in the ``shortcut'' approach one gives empirical evidence
that VA2--VA5 hold.
Of course, if one is able to provide compelling evidence of both types,
so much the better.

\apasection{Analysis of Losada (1999)}

We begin this section by summarizing Losada's (1999) experimental setup
and briefly pointing out some serious deficiencies therein.
These deficiencies would suffice, by themselves, to render
Losada's empirical work of limited or no scientific value;
but they are of only minor importance compared to the principal
abuse in Losada's article, which is the purported application
of the Lorenz equations to describe the changes in human emotions
over time.  In the preceding section we listed five criteria,
VA1--VA5, {\em all}\/ of which must be met if an application of
differential equations is to be valid. 
Here we shall demonstrate that
{\em none}\/ of these criteria have been met by Losada's work,
except arguably a small part of VA1.
We shall also demonstrate that the conditions for validity
of the ``shortcut'' approach have not been met either.

\apasubsection{Losada's experiments}

Losada (1999) described a series of experiments that he
conducted during the 1990s while he was the director of a
laboratory (``Capture Lab'') run by the U.S.~computer services
corporation EDS.  Within this laboratory, 60 business teams
were studied during meetings, while observers positioned behind
one-way mirrors analyzed and coded their verbal communications;
some details of the procedures used were described by Losada
and Markovitch (1990).  These coded data form the entire
empirical basis of all the subsequent work allegedly leading to the
theoretical derivation of the critical minimum positivity ratio
using a nonlinear-dynamics model.

Losada's (1999) article
followed few of the conventions that would
normally be expected from a piece of scholarship published
in a scientific journal.  It presented very little primary
data, and the experimental design, construction of models,
and interpretation of results were made with little or
no justification.  Indeed, some aspects of the methods and
results of the Capture Lab experiments were described for the
first time only in subsequent articles or books,
while many other crucial aspects remain obscure to this day.
Many important details of the ``business teams,'' and of their
meetings that were observed in the laboratory setting, are not clear.
Losada (1999, p.~188) stated
that, because of capacity constraints in the Capture Lab,
each of the 60 observed teams was composed of exactly eight members;
but he did not say whether any otherwise-eligible teams
had a ``real-world'' membership larger (or smaller) than eight,
and if so, whether they were truncated (or expanded)
for the experiment or simply excluded from participation.
Losada also
did not indicate how long the meetings lasted,
though Fredrickson (2009, p.~123) finally reported, informally,
that these were ``hour-long'' meetings.
Not even the most basic demographic information, such as the
sex ratio or the mean ages of the participants, was provided.
Above all,
neither Losada (1999) nor
either of the two later articles
discussed any possible effects on the participants of being observed and recorded
during their workplace meetings.  We consider these to be
substantial omissions for an article describing an empirical
experiment in psychology, especially one whose results have
been used with such portentous theoretical consequences.

The ``speech acts'' of the members of the business teams
during their meetings were coded according to three bipolar dimensions:
``positivity--negativity,'' ``other--self,'' and ``inquiry--advocacy''
(Losada, 1999, p.~181).
It is not clear whether each speech act was coded according
to all three dimensions, or to only one.

Having rated teams as ``high performance,'' ``medium performance,''
or ``low performance'' based on business indicators,
Losada (1999, p.~180) next proceeded to analyze the teams'
``degree of connectivity'' or ``nexi index,''
which he defined as ``the number of cross-correlations
significant at the .001 level or better
that were obtained through the time series analysis of the data
generated by coding speech acts at the Capture Lab.''
This is a crucial piece of information because Losada went on to use its
value directly as the control parameter
--- which he called $c$, and which Lorenz (1963) called $r$ ---
in the Lorenz equations.
Losada gave no precise definition, however,
of what he meant by ``number of cross-correlations,'' nor did he specify
what statistical test he was using to define the significance
level, nor why he chose the particular level of 0.001;
note that any other choice would have given a different value for
the Lorenz control parameter $c$.
Losada (1999, p.~180) did say vaguely that these cross-correlations
``represent sustained couplings or matching patterns
of interlocked behaviors among participants throughout the
whole meeting.''
{}From remarks made later in the article (p.~188)
as well as slightly more detailed explanations
given in Losada and Heaphy (2004, pp.~747--748, 763)
and Fredrickson (2009, p.~124),
we tentatively infer that ``the number of cross-correlations''
might mean the number of ordered pairs $i,j$ of team members ($i \neq j$)
such that the cross-correlation function $C_{ij}(t)$ of their time series
meets certain unspecified criteria.\footnote{
   Losada and Heaphy (2004, p.~763, note 2)
   cited Losada, S\'anchez, and Noble's (1990)
   concept of a ``group interaction diagram''
   --- a sort of directed graph in which the nodes are the group members
   and the arrows indicate cross-correlations that are nonzero and
   statistically significant under some unspecified statistical tests ---
   as providing ``a graphical representation of nexi''.
   This supports our tentative inference about the
   probable intended meaning of Losada's (1999) term
   ``number of cross-correlations''.
}
The maximum possible number of such ``nexi'' for an eight-person group
would therefore be $8 \times 7 = 56$.\footnote{
   The alternative guess that ``nexi'' refer to {\em unordered}\/ pairs
   of team members is contradicted by the fact that
   the maximum possible number of such ``nexi'' for an eight-person group
   would be 28, while Losada's high-performance teams had an average
   nexi index of 32.
   Note, however, that if ``nexi'' refer to {\em ordered}\/ pairs
   of team members, then the criteria defining a nexus
   --- whatever they may be, as Losada (1999) has not told us ---
   would have to be time-asymmetric, since $C_{ij}(t) = C_{ji}(-t)$;
   otherwise we would, de facto, be using unordered pairs after all.
}
The rounded average nexi for
high-performance, medium-performance and low-performance teams
were 32, 22, and 18, respectively (Losada, 1999, p.~180).

\apasubsection{Losada's use of differential equations}

Having examined some of the deficiencies
in Losada's (1999) description of the empirical part of his research,
we now move on to the fundamental
problem
in the article:  namely, the use of differential equations to model
the time evolution of human emotions in general,
or interactions within business teams in particular.
We organize our analysis according to the criteria VA1--VA5
(set forth previously) for the valid application of differential equations.
This analysis divides naturally into two parts:
the general use of differential equations (VA1--VA4),
and the specific use of the Lorenz system (VA5).

\apasubsubsection{VA1}
Losada (1999, p.~182) stated that the variables $X,Y,Z$ represent, respectively,
``inquiry--advocacy,'' ``other--self,'' and ``emotional space.''
Earlier (p.~181) he defined ``emotional space'' as
``the ratio of positivity to negativity.''
But he did not indicate anywhere in the article
whether $X$ and $Y$ are defined as {\em ratios}\/
of ``inquiry'' to ``advocacy''  and ``other'' to ``self,'' respectively,
or as {\em differences}\/.
(In fact, {\em either}\/ option leads to severe difficulties
invalidating his use of the Lorenz equations, as discussed below.)

Even more importantly, Losada (1999) gave no explanation of how the ``speech acts,''
which are discrete events occurring at discrete moments of time,
are to be converted into smoothly varying functions $X(t), Y(t), Z(t)$
of a continuous time variable $t$.
Arguably this {\em could}\/ be done, at least approximately,
by subdividing the one-hour session into time intervals $\Delta t$
that are both
(a) long enough so that each interval contains a large number of ``speech acts''
(so that the discrete ratios or differences could be approximated in a sensible
way by a continuous variable $X$, $Y$, or $Z$), and
(b) short enough so that the change in $X,Y,Z$ from one interval to the next
is small (so that the discrete time variable indexing the intervals could be
approximated in a sensible way by a continuous variable $t$,
with $X,Y,Z$ varying reasonably smoothly as a function of $t$).
It is at least conceivable that these two conditions could be
simultaneously fulfilled, though we think it unlikely.\footnote{
   Five years later, Losada and Heaphy (2004, p.~745) finally revealed that
   ``data generated by the coders were later aggregated in one-minute
   intervals.''  They indicated that ``time series analyses, including
   the auto-correlation and cross-correlation function, were performed on
   these aggregated data'';  but they did not explain whether (or how) these
   aggregated data were also used to construct the smoothly varying functions
   $X(t), Y(t), Z(t)$ whose time evolution is allegedly modeled
   by the Lorenz system, nor did they present any information whatsoever
   concerning these time series other than their means (p.~747, Table~1).
   Curiously, Table~1 presents the means for the {\em ratios}\/
   in all three categories,
   although Losada and Heaphy (2004, p.~754) later
  implied
  that
   $X$ and $Y$ might be defined via {\em differences}\/
   rather than ratios (see the discussion below).
}

The key problem, however, is not simply that Losada (1999) failed to {\em explain}\/
any of this to his reader.
The problem is, rather, that there is not
the slightest evidence that Losada {\em did}\/ any of this.
Nowhere in
Losada (1999) --- or, for that matter, in the two subsequent articles ---
is there any description of empirical data
that have the form of smoothly varying functions $X(t), Y(t), Z(t)$
of a continuous time variable $t$.
Indeed, none of the three articles evinces any awareness
that a phenomenon describable by a differential equation
must involve quantities that vary smoothly as a function of time.

We conclude that Losada (1999) has
failed to satisfy the
most basic and elementary requirement VA1 for the applicability of
differential equations.
This alone would invalidate his work.

\apasubsubsection{VA2}
Losada (1999) did not give any arguments to support the idea that
his variables $X,Y,Z$ {\em evolve by themselves}\/;
the question was not even considered.
A priori it seems implausible that these three variables
should evolve autonomously, without significant influence from
other variables that might affect the emotional state of
a business meeting.

\apasubsubsection{VA3}
Losada (1999) did not give any arguments to support the idea that
his variables $X,Y,Z$ {\em evolve deterministically}\/;
once again, the question was not even considered.
A priori this hypothesis seems even more implausible than the preceding one.

\apasubsubsection{VA4}
Losada (1999) did not give any arguments to support the idea that
his variables $X,Y,Z$ {\em evolve according to a differential equation}\/,
that is, that the rate of change at each moment of time
depends only on the values of the variables at {\em that same}\/ moment of time
(and not on past values).
This too is a priori implausible
as it is tantamount to assuming that the participants in the meeting have no memory.
It is, of course, conceivable that change in emotions is dominated
(at least under some circumstances) by the current state of emotions,
so that the effect of memory is small;
if this were the case {\em and}\/ VA1--VA3 were also to hold,
then VA4 might be justified as a reasonable approximation.
But Losada did not even raise this possibility,
much less give any theoretical or empirical evidence to support it.

\apasubsubsection{VA5}
Finally, let us suppose,
for the sake of argument and contrary to
what we have just shown,
that Losada (1999) had adequately fulfilled criteria VA1--VA4,
that is, that he had precisely defined all his variables and had demonstrated
that their time evolution should be governed,
at least to a good approximation, by
a deterministic first-order differential equation.
The question would still remain:
What reasons did
Losada (1999)
give for supposing that the governing
equation should be the Lorenz system?

To see the type of reasons that are offered,
let us quote in full Losada's
(1999)
``derivation'' of his first equation,
which corresponds to Equation~\ref{eq.lorenz.c} above:
\begin{quote}
\small
Thinking about the model that would generate time series
that would match the general characteristics of the actual time series
observed at the Capture Lab [this issue will be discussed below],
it was clear that it had to include nonlinear terms representing the
dynamical interaction among the observed behaviors.
One such interaction is that between inquiry-advocacy and other-self.
If I call the first $X$ and the second $Y$, their interaction should be
represented by the product $XY$, which is a nonlinear term.
I also knew from my observations at the lab,
that this interaction should be a factor in the rate of change
driving emotional space (which I will call $Z$).
In addition, I would need a scaling parameter for $Z$.
Consequently, the rate of change of $Z$ should be written as
$$
   {dZ \over dt}  \;=\;  XY \,-\, aZ  \;,
$$
where $a$ is a scaling parameter that would be held constant.
(Losada, 1999, p.~182)
\end{quote}
Losada
(1999)
gave no explanation of why
the ``interaction'' between $X$ and $Y$, and only these,
should be significant in determining the rate of change of $Z$
(``positivity/negativity'').
It would seem equally plausible that the rate of change of $Z$
might be affected by the ``interaction'' of $X$ and $Z$,
or of $Y$ and $Z$, or of all three variables together.
Nor is it explained why this ``interaction'' should consist of
multiplying their values together;
indeed, if $X$ and $Y$ are dimensionless quantities
(as they must be for the Lorenz system to be valid),
any mathematical operation could be used to combine them, such
as the square root of the sum of their squares, or the difference of
their exponentials, or the product of their cube roots.
Finally, it is not explained why the rate of change of $Z$
should contain a term linear in $Z$ but not one in $X$ or $Y$.

Similar considerations apply to Losada's (1999, p.~182) subsequent
``derivation'' of every term in his three equations.
The principal attribute that all of the choices made
by Losada have in common is that they correspond to terms
in Lorenz's system of differential equations.
The reader is left with the feeling of having watched a video clip
of a Rubik's Cube being miraculously solved in five seconds,
only for it to be revealed at the end that what was filmed
was an ordered cube being scrambled, with the whole process
then being played back in reverse.\footnote{
   Luoma, H\"am\"al\"ainen,
  and
  Saarinen (2008)
   made considerable effort to connect Losada's (1999) work
   with their own concept of Systems Intelligence;
   even so, they felt compelled to observe (p.~760) that
   ``there is no further discussion [of] why these [Lorenz]
     equations should describe the dynamics of the
     coded observations of verbal communication.
     Only very limited explanations are given about
     the modelling process and the meaning and
     interpretation of its parameters (see Losada,
     1999, p.~182). Thus, the reasoning behind the
     model equations remains unclear to the reader.''
}

A final problem is posed by Losada's (1999) lack of clarity
as to whether his variables $X$ and $Y$ are defined as {\em ratios}\/
of ``inquiry/advocacy''  and ``other/self,'' respectively,
or as {\em differences}\/.
Either option leads to fundamental contradictions with the Lorenz system.
If $X$ and $Y$ are ratios of speech acts
(as the analogy with $Z$ would suggest),
then they are dimensionless and obviously nonnegative.
If $X$ and $Y$ are differences of speech acts
--- as Losada and Heaphy (2004, p.~754) seemed to imply when discussing
the values plotted on the x-axis of their Figure~5 ---
then they are dimensionful
(they would be rates of speech acts per unit time,
so that their numerical values would depend on the choice of the unit of time)
and take both positive and negative values
(depending, e.g., on whether inquiry predominates over advocacy or vice versa).
In the Lorenz system, by contrast,
$X$ and $Y$ are dimensionless and take both positive and negative values
(as Losada's Figures~1--6 clearly show).
Therefore, no matter how Losada's
(1999)
unclarity about the definition of $X$ and $Y$
were to be resolved, those variables would not have the properties
assumed in the Lorenz system.

\apasubsection{The alleged match between Losada's empirical data and his
   mathematical model}

Losada (1999, p.~183) asserted that ``The time series generated by this model
matched all the general characteristics of the time series observed at the lab
for each team performance level.''
However, he
did not provide {\em any}\/ empirical data
showing the time series $X(t), Y(t), Z(t)$ allegedly produced by his lab data;
indeed, as discussed earlier, he failed to explain, even in the vaguest terms,
how such smoothly varying functions were supposedly constructed
from his raw data on ``speech acts.''
Nor did he explicate the nebulous phrase ``general characteristics.''
Losada's
(1999)
claim that his model matches his data must therefore be taken entirely
on faith.

For what it is worth, it seems a priori unlikely that
the emotions of
business teams
composed of
normal
individuals
would oscillate
wildly and continuously throughout
a one-hour meeting in the manner of the Lorenz attractor.
For instance, Losada's (1999) Figures~1 and 2 allegedly show the ratio
of positivity to negativity (what Losada called ``emotional space'')
for high-performance teams fluctuating repeatedly between a minimum of about 10
and a maximum of about 50.  Both numbers are thoroughly implausible,
as is the factor-of-5 swing between them.

It follows that Losada (1999) has also failed to meet
the minimal conditions for validity of the ``shortcut'' approach
to modeling using differential equations,
for at least four reasons (any one of which would be fatal):
the variables are not clearly defined (VA1);
the purported empirical verification is weak or nonexistent;
it is a priori implausible that the data could match the model
even qualitatively, much less quantitatively (VA5);
and it is a priori implausible (as noted earlier) that
even VA2--VA4 could hold for the system under study.

What Losada (1999) did in Section~5 (pp.~183--188) of his article
was simply to run some computer simulations
of the Lorenz equations at parameters
$\sigma = 10$, $b = 8/3$, and $r = 32,22,18$.
To the results of these purely mathematical simulations,
shown in his Figures~1--9,
Losada then
appended the
words
``inquiry--advocacy,'' ``other--self,'' ``emotional space,''
and ``connectivity'' (or ``number of nexi'')
to the mathematical quantities $X$, $Y$, $Z$, and $r$, respectively.
He referred to the simulations at $r = 32,22,18$
as describing the trajectories of
``high performance teams'' (p.~183),
``medium performance teams'' (pp.~187--188),
and ``low performance teams'' (pp.~184--185),
respectively;
but he presented no data
to support this purported identification.

Let us also note
that both $r=22$ and $r=18$ lie in the region 
$r < r_{\rm A} \approx 24.06$ (see Footnote~\ref{footnote_rA})
where the solution of the Lorenz system
settles down, as $t \to\infty$, to a fixed point
at $X = Y = \pm \sqrt{b(r-1)}$, $Z = r-1$.
The results for ``medium performance teams'' and ``low performance teams''
shown in Losada's (1999) Figures 4--9 are thus transient behavior,
not indicative of the system's long-term state,
and reflective only of Losada's arbitrary (and unspecified) choices
of initial conditions and total run length.
In particular, Losada's assertion (p.~188) that the $r=22$ data
end up in a limit cycle
---
repeated in Losada and Heaphy (2004, pp.~751, 755, 762),
Fredrickson and Losada (2005, pp.~682, 683, 685)
and Fredrickson (2009, pp.~126, 129) ---
is incorrect.\footnote{
   Although dynamical-systems theory focuses principally on the system's
   long-time behavior, it should be stressed that the transient behavior
   is not necessarily without interest.
   Our goal in the present paragraph is merely to observe that a central thesis
   of the three subject articles ---
   namely, the parallel between the authors' concepts of
   low-, medium- and high-performance teams
   and the mathematical concepts (which indeed pertain to long-time behavior)
   of fixed points, limit cycles, and strange attractors ---
   is fatally flawed by (among many other things) an erroneous understanding
   of the behavior of the Lorenz system.
 \label{footnote_transient}
}

But let us once again put aside all the foregoing criticisms and suppose,
just for the sake of argument and contrary to all evidence,
that Losada (1999) had adequately demonstrated that the time evolution of emotions
within his teams' business meetings was governed by the Lorenz system.
There still arises the question of how the dimensionless parameters
$\sigma$, $b$, and $r$ in the Lorenz equations are to be chosen.
For $\sigma$ and $b$, Losada (p.~183) simply followed the arbitrary choice
made for purely illustrative purposes by Saltzman (1962) and Lorenz (1963),
and set $\sigma = 10$ and $b = 8/3$.
For the key parameter $r$, Losada (p.~182) set this equal to the number of nexi ($c$),
but without providing any justification for doing so.
In fact, choosing $r$ {\em equal to}\/ (rather than merely {\em dependent on}\/)
the number of nexi is
puzzling,
since $r$ is a universal control parameter in the Lorenz system,
while the number of nexi obviously depends on the size of the business team.
Consequently,
even if the ``results'' of the three subject articles were somehow to be
scientifically valid (and of course they are not),
they would apply at best to the dynamics of eight-person teams,
rather than being
universal truths about
human emotions.
\apasubsection{Conclusion}

One can only marvel at the astonishing coincidence that human
emotions should turn out to be governed by exactly the same
set of equations that were derived in a celebrated article
several decades ago
as a deliberately simplified model of convection in fluids,
and whose solutions
happen to have visually appealing properties.
An alternative explanation --- and, frankly, the one that
appears most plausible to us --- is that the entire process
of ``derivation'' of the Lorenz equations has been contrived
to demonstrate an imagined fit between some rather
limited empirical data and the scientifically impressive world
of nonlinear dynamics.

But Losada (1999) goes farther:

\begin{quote}
\small
An interesting observation that highlights the usefulness
of fluid dynamics concepts to describe human interaction
arises from the fact that Lorenz chose the Rayleigh number
as a critical control parameter in his model.  This number
represents the ratio of buoyancy to viscosity in fluids.
A salient characteristic of my observations of teams at the
Capture Lab was that high performance teams operated in a
buoyant atmosphere created by the expansive emotional space
in which they interacted and that allowed them to easily
connect with one another.  Low performance teams could be
characterized as being stuck in a viscous atmosphere highly
resistant to flow, created by the restrictive emotional space
in which they operated and which made very difficult for them
to connect with one another;
hence, their nexi were much lower
than the nexi for high performance teams.
(Losada, 1999, p.~183)
\end{quote}
Let us pass quickly over
the notion that Lorenz simply ``chose''
to use the Rayleigh number in his fluid-dynamics equations
(just as Einstein perhaps ``chose'' to use the speed of light
in his equation $E=mc^2$?)\@
as well as the minor technical error in the second sentence of this
paragraph (the ratio of buoyancy to viscosity in fluids is
the Grashof number, not the Rayleigh number).
Instead, we
invite the reader to contemplate
the implications of the third and fourth sentences.
They appear to assert that the predictive use of differential equations
abstracted from a domain of the natural sciences
to describe human interactions
can be justified on the basis of the {\em linguistic}\/ similarity
between elements of the technical vocabulary of that scientific
domain and the adjectives used metaphorically by a particular
observer to describe those human interactions.  If true, this
would have remarkable implications for the social sciences.
One could describe a team's interactions as ``sparky'' and
confidently predict that their emotions would be subject to
the same laws that govern the dielectric breakdown of air
under the influence of an electric field.  Alternatively, the
interactions of a team of researchers whose journal articles
are characterized by ``smoke and mirrors'' could be modeled
using the physics of airborne particulate combustion residues,
combined in some way with classical optics.

\apasection{Analysis of Losada and Heaphy (2004)}

Losada and Heaphy (2004) claimed
to extend Losada's (1999) ``findings''
by demonstrating that the latter's construct of ``connectivity'' is,
in fact, directly arithmetically related to the positivity-negativity (P/N) ratio.
In this section we shall try, as best we can,
to follow Losada and Heaphy's
(2004)
reasoning.
For the purposes of this exercise, we invite the reader to suppose,
for the sake of argument and contrary
to everything that we have just demonstrated,
that all
of Losada's
(1999)
``results''
are entirely correct; in this way we can highlight
(within the limits of the space available to us)
the independent deficiencies of logic
in
Losada and Heaphy's
article.\footnote{
Let us warn the reader that, confusingly,
Losada and Heaphy (2004) have
interchanged the names of the variables $Y$ and $Z$
compared to the convention employed by Lorenz (1963) and Losada (1999).
For clarity we will continue to use Lorenz's (1963) conventions
as in Equations~\ref{eq.lorenz} above;
this must be borne in mind when comparing our formulae
with those of Losada and Heaphy (2004).
   \label{footnote_LH_notation_change}
}

\apasubsection{The redefinition of ``emotional space''}

Recall that Losada (1999, p.~181) defined ``emotional space''
as ``the ratio of positivity to negativity.''
But, as noted above, this definition is blatantly incompatible
with Losada's (1999, p.~182) identification of ``emotional space''
with the Lorenz variable $Z$,
since the latter fluctuates (when $\sigma=10$, $b=8/3$, $r=32$)
between approximately 10 and 50.
Perhaps conscious of this, Losada and Heaphy
(2004)
opted to change (entirely arbitrarily) the definition of ``emotional space'':
\begin{quote}
\small
We know that emotional space is generated by the P/N ratio.
The scale of the y-axis
[note that Lorenz (1963) and Losada (1999) called this $Z$]
does not represent directly the P/N ratio, but the outcome of the initial
value (16) entered into the equation to eliminate the transient (this is a standard
procedure in nonlinear dynamics and in modeling in general) and the multiplication
by the constant 8/3 (a constant used in all Lorenz system models). By
introducing the initial value and multiplying by a constant we are creating an initial
emotional space that will stay there increased or decreased by the P/N ratio.
(Losada \& Heaphy, 2004, p.~754) 
\end{quote}
This paragraph is somewhat confusing, but let us do our best to decipher it.

In the first sentence,
Losada and Heaphy (2004) have replaced Losada's (1999, p.~181) ``equal to''
with a much vaguer ``generated by'';
and in the first half of the second sentence they have explicitly denied
Losada's definition (but without admitting this forthrightly).
What comes next is, however,
not merely incorrect but
in fact the exact opposite of the truth.
The initial conditions (such as Losada's apparent choice $Z_0 = 16$)
are essentially {\em irrelevant}\/ to determining the long-time behavior
of the system.\footnote{
   At least when there is a unique stable attractor, as is the case for $r=32$.
   For $r=22,18$ there are {\em two}\/ stable fixed points,
   at $X = Y = \pm \sqrt{b(r-1)}$, $Z = r-1$,
   and the initial conditions determine which one of them is reached
   as $t \to\infty$.
}
Depending on what initial condition is chosen,
one may have to wait a longer or shorter time (the ``transient'') before reaching
the true long-time behavior (e.g., the Lorenz attractor);
but that ultimate behavior (including what Losada and Heaphy call
``the scale of the y-axis'') is determined solely by the
parameters $\sigma$, $b$, and $r$
in the Lorenz system (Equations~\ref{eq.lorenz} above),
{\em not}\/ by the initial condition.
Finally, the remainder of this paragraph
--- that is, the comment regarding ``multiplication by the constant 8/3''
as well as the entire final sentence ---
obeys no logic that we are able to discern.
But it does lead to the apparently desired result:
namely, ``emotional space'' is no longer {\em equal to}\/ the P/N
ratio --- despite this being Losada's (1999, p.~181) own definition of the term ``emotional space'' ---
but is
apparently
related to it by some as-yet-unspecified formula
involving addition (or subtraction) and multiplication.

Having redefined ``emotional space'' as something other than the P/N ratio,
Losada and Heaphy's (2004) next task was to
``address the question of whether emotional space is linked to connectivity
and how emotional space is specifically related to the positivity to
negativity ratio'' (p.~755).
They did this in several
steps, as described in the following subsections.

\apasubsection{Linking ``emotional space'' and ``connectivity''}

Losada and Heaphy's (2004) first step was to draw attention to the
``blank space approximately in the middle of the attractor in each of its wings''
(p.~755), which they called the ``focus'' of the attractor.
These ``foci'' are nothing other than the unstable (for $r > r_{\rm crit}$)
fixed points of the Lorenz equations, which as we saw earlier
are located at $X = Y = \pm \sqrt{b(r-1)}$, $Z = r-1$.
Losada and Heaphy did not
mention this fact,
but they did observe empirically that
the foci are located at $Z = r-1$ for their three simulations,
which they rewrote as $E = c-1$,
``where $E$ is emotional space, and $c$ is connectivity
(represented by the number of nexi)'' (pp.~755--756).

\apasubsection{Linking ``emotional space'' and the positivity-negativity ratio}

Losada and Heaphy's (2004) next step was to link ``emotional space'' ($E$)
to the P/N ratio.  They began by asserting that
\begin{quote}
\small
When running the ML [meta learning] model initial values as well as scaling
constants must be assigned. The initial values eliminate
transients, which represent features of the model that
are neither essential nor lasting. The initial value for
positivity/negativity is 16. The constants are used to scale
the data, namely to be able to see the dynamics more clearly.
(Losada \& Heaphy, 2004, p.~757) 
\end{quote}
This paragraph repeats the
incorrect statements about the initial conditions,
which as we have seen play {\em no}\/ role in determining the long-time behavior
of the system (namely, the features that are ``essential'' and
``lasting'').\footnote{
   Let us observe once again (see Footnote~\ref{footnote_transient} above)
   that transient behavior is not necessarily without interest.
   But Losada and Heaphy (2004, p.~757) have here stressed
   that {\em their own}\/ interest was focused on the long-time behavior,
   not the transients.
 \label{footnote_transient_2}
}
Therefore, {\em pace}\/ Losada and
Heaphy,
a valid equation concerning the long-time behavior {\em cannot}\/
involve the initial conditions.
Furthermore, the constants $\sigma$, $b$, and $r$ are simply parameters in
the Lorenz system;
there is no
reason why they should be ``used to scale the data.''
But no matter:
after drawing attention to the constant $b=8/3$,
Losada and Heaphy
completed their ``reasoning'' as follows:
\begin{quote}
\small
With this background information, we can now calculate the
P/N ratio. To derive the P/N ratio from the attractor's foci,
we subtract the initial value and multiply it by the inverse of
the scaling constant (0.375). For example, for high performance
teams, we start with 31, subtract 16, and multiply by 0.375. The
result is 5.625, which is very close to 5.614, the result
obtained by looking at the original time series data. We can
now introduce the equation that allows us to calculate the
positivity to negativity ratio (P/N) from emotional space (E):
\vspace*{-2.5mm}
$$
   {\rm P/N}  \;=\;  (E-i) \, b^{-1}
\vspace*{-2.5mm}
$$
where $E$ is emotional space, $i$ is the initial value of the
positivity/negativity state variable (equal to 16),
and $b^{-1}$ is the inverse scaling constant (equal to 0.375).
If we apply this formula to the $E$ numbers for medium (21)
and low-performance teams (17), we obtain results that
are equally close to the ones obtained by looking directly
at the time series data, thus further validating the ML model \ldots\ 
(Losada \& Heaphy, 2004, p.~757) 
\end{quote}
Indeed, the correspondence between data and theory is astounding
(Losada \& Heaphy, 2004, p.~758, Table~2):
\begin{center}
\vspace*{-2mm}
\small
{\bf Positivity/Negativity Ratios from Time Series and Model}
\begin{tabular}{lcc}
& {\em Time Series Data}\/ &  {\em Model Data}\/   \\[-1mm]
High-performance teams    &  5.614  &  5.625  \\[-3.5mm]
Medium-performance teams  &  1.855  &  1.875  \\[-3.5mm]
Low-performance teams     &  0.363  &  0.375
\end{tabular}
\end{center}
And yet, the manipulations leading to the key equation
${\rm P/N} = (E-i) \, b^{-1}$ are completely arbitrary.
This whole exercise seems to be a post hoc ``justification,''
with the artificial offset and scaling being applied to the model
merely to allow it to produce ``predictions'' that resemble the observed data.

\apasubsection{Linking ``connectivity'' and the positivity-negativity ratio}

Combining the ``results'' of the preceding two sections,
Losada and Heaphy (2004) deduced the desired connection between
``connectivity'' and the P/N ratio:
\vspace*{-8.5mm}
\begin{quote}
\small
$$
   {\rm P/N}  \;=\;  (c-i-1) \, b^{-1}
\vspace*{-2mm}
$$
where P/N is the ratio of positivity to negativity, $c$ is connectivity defined by the
number of nexi, $i$ is the initial value of the positivity/negativity state variable
and $b^{-1}$ is the inverse scaling constant.
(Losada \& Heaphy, 2004, p.~758) 
\end{quote}
This final step is, of course, perfectly correct algebra,
if one grants
what has come before.

\apasubsection{Conclusion}

We confess that we have been unable to identify any mathematically or psychologically
meaningful reasoning or analysis in Losada and Heaphy's (2004) derivation
of their main result, ${\rm P/N} = (c-i-1) b^{-1}$.
As in Losada (1999), there are plenty of mathematical formulae --- albeit here elementary algebra
rather than nonlinear differential equations ---
but their only function, as far as we can tell, is to create, without any apparent justification,
an equation that purports to describe a relationship between the P/N ratio and ``connectivity''
and that happens to provide a good straight-line fit to three data points.

\apasection{Analysis of Fredrickson and Losada (2005)}

Fredrickson and Losada's (2005) central claim --- and their
key innovation beyond
Losada (1999) and Losada and Heaphy (2004) ---
was the purported existence of a critical minimum positivity ratio
value of
2.9013.
Fredrickson and Losada took over the ``results'' of the two earlier articles as
{\em faits accomplis}\/,
with little or no explanation of the
logic by which they were allegedly deduced.
This seems to us a rather grave omission in an article
that makes such extraordinary theoretical claims.
Nevertheless, we once again invite the reader
to assume for the sake of argument that
the two preceding articles are correct in every detail;
in this way we can highlight the independent flaws
in Fredrickson and Losada (2005).

\apasubsection{Derivation of the critical minimum positivity ratio 2.9013}

Fredrickson and Losada (2005), after briefly recounting some studies suggesting
that ``high ratios of positive to negative affect would distinguish individuals
who flourish from those who do not'' (p.~680),
then went on to assert that their more radical contention that
``individuals or groups must meet or surpass a specific
positivity ratio to flourish'' is supported by ``a nonlinear dynamics
model empirically validated by Losada (1999)'' (p.~681).
After briefly summarizing Losada's (1999) experimental setup,
they purported to explain the logic of Losada's work as follows:
\begin{quote}
\small
Observation of the structural characteristics (i.e.,
amplitude, frequency, and phase) of the time series of the
empirical data for these three performance categories led
Losada to write a set of coupled differential equations to
match each of the structural characteristics of the empirical
time series. Table~1 presents these equations
[which are the Lorenz equations but with $Y$ and $Z$ interchanged].
Model-generated time series were subsequently matched to the empirical
time series by the inverse Fourier transform of the
cross-spectral density function, also known as the
{\em cross-correlation function}\/.
Goodness of fit between the mathematical
model and the empirical data was indicated by the
statistical probability of the cross-correlation function at $p < .01$.
(Fredrickson \& Losada, 2005, p.~681)
\end{quote}
The trouble is, no such statistical matching between the model-generated time series
and the empirical time series is anywhere even alluded to --- much less presented ---
in Losada (1999).
(Indeed, as noted earlier, no empirical data for the variables
$X,Y,Z$ are presented at all.)
Nor, for that matter, is any such matching mentioned anywhere
in the later article of Losada and Heaphy (2004).
The alleged empirical validation of Losada's mathematical model is
thus unsupported by anything in the three subject articles.
(Let us therefore put aside the otherwise crucial question of precisely what
 statistical procedures were employed to perform this matching, and whether
 they are valid.)

Fredrickson and Losada (2005) then went on to present the famous
``butterfly'' plots showing the trajectories of the Lorenz system
(p.~682, Figure~1).
They asserted that the three plots are, respectively,
``derived from the empirical time series of the flourishing,
high-performance teams'' (p.~681),
``derived from the empirical time series of the medium-performance teams'' (p.~681),
and ``derived from the empirical time series of the low-performance teams'' (p.~682)
--- they repeated this phrasing verbatim three times ---
but they provided no evidence
that these claims are true
(neither, as we saw, did Losada, 1999).
Figure~1 shows the results of
computer simulations of the Lorenz equations,
nothing more.

Finally, Fredrickson and Losada (2005) derived the ideal minimum positivity
ratio of 2.9013 by recalling (pp.~682--683) that
\begin{quote}
\small
Subsequent work on the model (Losada \& Heaphy, 2004)
revealed that the positivity ratio relates directly to
the control parameter by the equation
$P/N = (c - Y_0 - 1) b^{-1}\,$ \ldots\ 
Past mathematical work on Lorenz systems by Sparrow (1982)
and others (Fr\o{}yland \& Alfsen, 1984; Michielin \& Phillipson, 1997)
has established that when $r$, the control parameter in
the Lorenz model, reaches 24.7368, the trajectory in phase
space shows a chaotic attractor.  Losada (1999) established the
equivalence between his control parameter, $c$, and the Lorenzian
control parameter, $r$.  Using the above equation,
it is known that the positivity ratio equivalent to $r=24.7368$ is 2.9013.
\end{quote}
Fredrickson and Losada
did not explain where $r_{\rm crit}=24.7368$
comes from or exactly how it leads to $(P/N)_{\rm crit} = 2.9013$, but
we can fill in the logic of their derivation
(making explicit the missing hypotheses)
and learn something new
in the process.
The first step
is to accept uncritically
the main ``result'' of Losada and Heaphy (2004),
namely, $P/N = (c-i-1) b^{-1}$  (here
Fredrickson and Losada
have renamed $i$ as $Y_0$).
The second step is to accept that Losada (1999) ``established''
the
equivalence
of $c$ and $r$ (although in fact he merely declared it by fiat, with
curious
consequences that we have already noted).
We now set $c$ equal to the value
$r_{\rm crit} = \sigma (\sigma+b+3)/(\sigma-b-1)$
that constitutes the boundary between non-chaotic and chaotic behavior
in the Lorenz system.
(When $\sigma = 10$ and $b=8/3$, this yields $r_{\rm crit}= 470/19 \approx 24.7368$.)
Simple algebra then gives the final formula
\be
   (P/N)_{\rm crit}   \;=\;
   {\sigma (\sigma+b+3)  \over  b(\sigma-b-1)}
   \:-\:  {i+1 \over b}
   \;.
 \label{eq.master}
\ee
Specializing to $\sigma = 10$, $b=8/3$ and $i=16$, we obtain
\be
   (P/N)_{\rm crit}   \;=\;
   {441 \over 152}  \;=\;
   2.901 \overline{315789473684210526}
\ee
where $\overline{\phantom{222}}$ denotes an infinitely repeating decimal.
Fredrickson and Losada (2005) were therefore far too modest to claim
only five significant digits;
their critical positivity ratio is in fact an exact rational number!
Unfortunately, there is one final, yet crucial, flaw lurking here:
the values of $\sigma$, $b$, and (especially) $i$
plugged into Equation~\ref{eq.master}
are {\em totally arbitrary}\/, at least within wide limits;
so the predicted critical positivity ratio is
{\em totally arbitrary as well}\/.
Choose different values of the parameters $\sigma,b,i$
and one gets a completely different prediction for $(P/N)_{\rm crit}$.
Recall that Saltzman (1962) chose $\sigma = 10$
for illustrative purposes and purely for convenience;
then Lorenz (1963) and Losada (1999) followed him.
Were humans to have eight fingers on each hand instead of five, 
Saltzman, and in turn presumably Lorenz and Losada,
might well have chosen $\sigma = 16$ instead of $\sigma = 10$
--- which (with $b=8/3$) produces a very similar Lorenz attractor,
except that the borderline of chaos is now
$r_{\rm crit} = 1040/37 = 28.\overline{108}$,
and the predicted critical positivity ratio
(with $i=16$)
is $(P/N)_{\rm crit} = 1233/296 = 4.1655\overline{405}$.
Yet other values of $\sigma,b,i$
would yield still different predictions for $(P/N)_{\rm crit}$.

Thus, even if one were to accept for the sake of argument
that every single claim made in Losada (1999) and
Losada and Heaphy (2004) is
correct,
and even if one were to further accept that the Lorenz equations
provide a valid and universal way of modeling human emotions,
then the ideal minimum positivity ratio that Fredrickson and
Losada (2005) claimed to have derived from Losada's
``empirically validated'' nonlinear-dynamics model
would still be nothing more than an artifact of
the arbitrary choice of an illustratively convenient value
made by a
geophysicist in Hartford in 1962.

\apasubsection{Fredrickson and Losada's empirical study}

In view of the foregoing vitiation of Fredrickson and Losada's (2005) ``derivation''
of the ideal minimum positivity ratio 2.9013,
a deep analysis of the results from their empirical study would be superfluous.
The fact that ``flourishing'' college students exhibited higher average
positivity ratios (3.2) than those who were ``languishing'' (2.3)
should not come as a surprise;
there is nothing inherently implausible about the idea that people
with a higher ratio of positive to negative emotions might experience
better outcomes than those with a lower ratio.  But the suggestion
that people with a positivity ratio of 2.91
are in some discontinuous way significantly better off than those with a
ratio of 2.90, simply because this number has crossed some magic line,
is not supported by any evidence.

\apasubsection{An upper critical positivity ratio?}

Compared to the abuse of mathematics in the theoretical part of their article,
Fredrickson and Losada's (2005) study
of ``flourishing'' versus ``languishing'' college students does
at least provide some
empirical evidence
that a higher positivity ratio typically
corresponds to better outcomes than a lower one.
By contrast,
their
claim that there is
also a precise desirable upper limit to the positivity ratio does not
appear to be supported by any evidence whatsoever.
Their prediction (p.~684) of an upper limit
of 11.6346
is
based entirely on the purported correspondence between human emotions
and the Lorenz system, combined with a gross mathematical error concerning the latter.
More precisely, Fredrickson and Losada equated psychological ``flourishing''
with the ``complex dynamics'' of the chaotic Lorenz attractor (pp.~682, 684);
therefore, they
(apparently)
reasoned,
if there is a value of the control parameter $r$
beyond which chaos no longer occurs,
then the end of chaotic attraction means the end of flourishing
--- in this case, at a positivity-ratio value that the authors
``estimate'' to be 11.6346 (p.~684).
Assuming that they arrived at this value using the same calculation method
as for the ideal minimum positivity ratio, this corresponds to $r \approx 48.0256$.
However, it is not the case that chaotic attraction in the
Lorenz system (with $\sigma = 10$ and $b = 8/3$) disappears beyond this value of $r$.
Rather, within the regime $r > r_{\rm crit} \approx 24.7368$
there is an intricate pattern of windows of periodicity (many of them quite narrow)
alternating or mixed with chaotic behavior, until at least $r \approx 200$
(Sparrow, 1982, Chapters~4 and 5, see especially Figure~5.12);
in particular, one extremely narrow periodic window appears to lie at
$48.0259 \ltapprox r \ltapprox 48.0271$
(Fr\o{}yland \& Alfsen, 1984; Fang \& Hao, 1996).
Thus, if one were to take seriously the reasoning of Fredrickson and Losada,
there would exist a complicated sequence of ``undesirable'' positivity-ratio intervals
lying within the mostly-desirable range above 2.9013.

But it goes without saying that this prediction is moot,
because, as we have demonstrated throughout the present paper,
the alleged connection between human emotions
and the Lorenz equations is entirely fanciful.

\apasubsection{Conclusion}

Fredrickson and Losada (2005) in effect claimed --- on the basis of an
analysis of verbal statements made in a series of one-hour meetings
held in a laboratory setting
by business teams of exactly eight people,
combined with some solemn invocations of the Lorenz equations ---
to have discovered a universal truth about human emotions,
valid for individuals, couples, and groups of arbitrary size
and capable of being expressed numerically to five significant digits.
This claim --- which was presented with no qualification or discussion of
possible limits to its validity --- would, if verified, surely require much of
contemporary psychology and neuroscience to be rewritten;
purely on that basis we are surprised that, apparently,
no researchers have critically questioned this claim,
or the reasoning on which it was based, until now.

We do not here call into question the idea that positive emotions are more
likely to build resilience than negative emotions,
or that a higher positivity ratio is ordinarily more desirable than a lower one.
But to suggest that some form of discontinuity sets in
at some special value of the positivity ratio
--- especially one that is independent of all demographic and cultural factors ---
seems far-fetched.
We cannot, of course, prove that no such ``tipping point'' exists;
but we believe that we have adequately demonstrated here that even if it does,
Fredrickson and Losada's (2005) article
--- based on a series of erroneous and, for the most part,
completely illusory ``applications'' of mathematics ---
has not moved science any nearer to finding it.

Fredrickson and Losada (2005, p.~685) concluded their article
by observing modestly that
``Our discovery of the critical 2.9 positivity ratio may
represent a breakthrough.''
Would that it were so.
\apasection{Concluding remarks}

The process that has taken place in this trio of articles
was presciently foreseen four decades ago
by the sociologist Stanislav Andreski:
\begin{quote}
\small
    The recipe for authorship in this line of business
   is as simple as it is rewarding:
   just get hold of a textbook of mathematics,
   copy the less complicated parts,
   put in some references to the literature in one or two branches
   of the social studies without worrying unduly about whether
   the formulae which you wrote down have any bearing on the
   real human actions, and give your product a good-sounding title,
   which suggests that you have found a key to an exact science
   of collective behaviour.
(Andreski, 1972, pp.~129--130)
\end{quote}
To be sure, Andreski's acerbic description of pseudoscientific work in
the social sciences may seem an exaggeration, and in most cases it
probably is.  But as applied to the articles of Losada (1999),
Losada and Heaphy (2004), and Fredrickson and Losada (2005),
Andreski's portrayal is, alas, literally accurate.

Let us stress
that our concern here is with the objective properties of published texts,
not the subjective states of mind of the authors
(which might, however, be of interest to philosophers, such as Frankfurt, 2005).
We do not, for example, have an opinion about the degree to which excessive enthusiasm,
sincere self-deception, or other motivations may have influenced Losada and colleagues
when writing their articles.
Our only interest here is to bring the fundamental errors in this widely-cited
body of work to the attention of the scientific community, before its considerable
influence can do any further damage to the cause of science in general
and academic psychology in particular.

More generally,
one of our aims in writing this article has been to alert scholars
who may be considering the use of differential-equation models in their work
to the need to ensure that their application of differential equations
to a specific natural or social phenomenon has been adequately justified,
be it by theoretical arguments or by empirical evidence or both.
To this end, we modestly propose that our checklist of criteria VA1--VA5
for the valid application of differential equations might be a useful tool.
In any case, we anticipate that the publication of the present paper
may stimulate a lively debate on this subject.

We wish to conclude on an optimistic note.
The fundamental property of science,
that it self-corrects and recovers from errors,
is often touted --- and rightly so --- as a strength
that distinguishes it from pseudoscience.
Unfortunately, in psychology just as in other scientific disciplines,
things do not always work out that way,
at least in the short run.
We are therefore grateful for the opportunity to publish this critique;
we hope that other scholars will be encouraged to question,
in public fora such as this one,
other research that seems to require correction.

\begin{center} {\bf References} \end{center}
\vspace*{-3mm}

\small

\begin{hangparas}{.5in}{1}

Andreski, S. (1972).  {\em Social sciences as sorcery}\/.  London, UK: Andre Deutsch.

Bales, R. F.  (1950).  {\em Interaction process analysis: A method for the
   study of small groups}\/.  Cambridge, MA: Addison-Wesley.

Bradburn, N. M. (1969).  {\em The structure of psychological well-being}\/.
Chicago, IL: Aldine.

Fang, H.-P., \& Hao, B.-L. (1996).
Symbolic dynamics of the Lorenz equations.
{\em Chaos, Solitons \& Fractals, 7}\/(2), 217--246.
doi:10.1016/0960-0779(95)00046-1

Frankfurt, H. G. (2005).  {\em On bullshit}\/.
Princeton, NJ: Princeton University Press.

Fredrickson, B. L. (1998).  What good are positive emotions?
{\em Review of General Psychology, 2}\/(3), 300--319.
doi:10.1037/1089-2680.2.3.300

Fredrickson, B. L. (2001).  The role of positive emotions in positive psychology:
The broaden-and-build theory of positive emotions.
{\em American Psychologist, 56}\/(3), 218--226.
doi:10.1037/0003-066X.56.3.218

Fredrickson, B. L. (2004).  The broaden-and-build theory of positive emotions.
{\em Philosophical Transactions of the Royal Society of London,
 Series B: Biological Sciences, 359}\/(1449), 1367--1377.
doi:10.1098/rstb.2004.1512

Fredrickson, B. L. (2009).
{\em Positivity: Groundbreaking research reveals how to embrace the hidden strength
of positive emotions, overcome negativity, and thrive}\/.
New York, NY: Crown.
Published in paperback as 
{\em Positivity: Top-notch research reveals the 3-to-1 ratio
that will change your life}\/.

Fredrickson, B. L., \& Kurtz, L. E. (2011).
Cultivating positive emotions to enhance human flourishing.
In S. I. Donaldson, M. Csikszentmihalyi, \& J. Nakamura (Eds.),
{\em Applied positive psychology: Improving everyday life, health, schools, work,
and society}\/ (pp.~35--48).  New York, NY: Routledge.

Fredrickson, B. L., \& Losada, M.F. (2005).
Positive affect and the complex dynamics of human flourishing.
{\em American Psychologist, 60}\/(7), 678--686.
doi:10.1037/0003-066X.60.7.678

Fr\o{}yland, J., \& Alfsen, K. H. (1984).
Lyapunov-exponent spectra for the Lorenz model.
{\em Physical Review A, 29}\/(5), 2928--2931.
doi:10.1103/PhysRevA.29.2928

Hilborn, R. C. (2000).  {\em Chaos and nonlinear dynamics: An introduction for
 scientists and engineers}\/.  New York, NY: Oxford University Press.

Kellert, S. H. (1995).  When is the economy not like the weather? The problem of
extending chaos theory to the social sciences.
In A. Albert (Ed.), {\em Chaos and society}\/ (pp.~35--47).  
Amsterdam, Netherlands: IOS Press.

Lorenz, E. N. (1963).  Deterministic nonperiodic flow.
{\em Journal of the Atmospheric Sciences, 20}\/(2), 130--141.
doi:10.1175/1520-0469(1963)020$<$0130:DNF$>$2.0.CO;2

Lorenz, E. N. (1993).  {\em The essence of chaos}\/.
Seattle, WA: University of Washington Press.

Losada, M. (1999).  The complex dynamics of high performance teams.
{\em Mathematical and Computer Modelling, 30}\/(9--10), 179--192.
doi:10.1016/S0895-7177(99)00189-2

Losada, M., \& Heaphy, E. (2004).
The role of positivity and connectivity in the performance of business teams:
A nonlinear dynamics model.
{\em American Behavioral Scientist, 47}\/(6), 740--765.
doi:10.1177/0002764203260208

Losada, M., \& Markovitch, S. (1990).
GroupAnalyzer: A system for dynamic analysis of group interaction.
In {\em Proceedings of the 23rd Hawaii international conference on system sciences}\/
(pp.~101--110).  Los Alamitos, CA: IEEE Computer Society Press.
doi:10.1109/HICSS.1990.205245
Losada, M., S\'anchez, P., \& Noble, E. E.  (1990).
Collaborative technology and group process feedback: Their impact on
interactive sequences in meetings.
In {\em CSCW '90 --- Proceedings of the 1990 ACM conference on
  computer-supported cooperative work}\/
(pp.~53--64).  New York, NY: Association for Computing Machinery.
doi:10.1145/99332.99341

Luoma, J., H\"am\"al\"ainen, R. P., \& Saarinen, E. (2008).
Perspectives on team dynamics: Meta learning and systems intelligence.
{\em Systems Research and Behavioral Science, 25}\/(6), 757--767.
doi:10.1002/sres.905

Michielin, O., \& Phillipson, P. E. (1997).
   Map dynamics study of the Lorenz equations.
   {\em International Journal of Bifurcation and Chaos, 7}\/, 373--382.
   doi:10.1142/S0218127497000248

Ruelle, D. (1994).  Where can one hope to profitably apply the ideas of chaos?
{\em Physics Today, 47}\/(7), 24--30.
doi:10.1063/1.881395

Saltzman, B. (1962).  Finite amplitude free convection as an initial value problem---I.
{\em Journal of the Atmospheric Sciences, 19}\/(4), 329--341.  
doi:10.1175/1520-0469(1962)019$<$0329:FAFCAA$>$2.0.CO;2

Seligman, M. E. P. (2011a).
{\em Flourish: A visionary new understanding of happiness and well-being}\/.
New York, NY: Free Press.

Seligman, M. E. P. (2011b, July 6).
{\em Lecture to the Royal Society for the Arts}\/ [Video file].
Retrieved from
http://www.thersa.org/events/video/vision-videos/martin-seligman.
See 02:55 -- 03:50.

Sokal, A., \& Bricmont, J. (1998).
{\em Fashionable nonsense: Postmodern intellectuals' abuse of science}\/.
New York, NY: Picador USA.
Published in the UK as
{\em Intellectual impostures: Postmodern philosophers' abuse of science}\/.
London, UK: Profile Books. 

Sparrow, C. (1982).
{\em The Lorenz equations: Bifurcations, chaos, and strange attractors}\/.
New York, NY: Springer-Verlag.

Stewart, I. (1997).  {\em Does God play dice?: The new mathematics of chaos}\/.
London, UK: Penguin.

Strogatz, S. H. (1994).  {\em Nonlinear dynamics and chaos: With applications to
physics, biology, chemistry, and engineering}\/.
Reading, MA: Addison-Wesley.

Williams, G. P. (1997).  {\em Chaos theory tamed}\/.
Washington, DC: Joseph Henry Press.

\end{hangparas}

\end{document}